\documentclass[a4paper,fleqn,usenatbib]{mnras}

\usepackage{newtxtext,newtxmath}

\usepackage[T1]{fontenc}
\usepackage{ae,aecompl}

\usepackage{graphicx}	
\usepackage{amsmath}	
\usepackage{amssymb}	
\usepackage[dvipsnames]{xcolor}
\usepackage{units}
\usepackage{textcomp} 

\title{Constraints on the structure of hot exozodiacal dust belts}

\author[F. Kirchschlager et al.
]{
Florian Kirchschlager$^{1}$\thanks{E-mail: kirchschlager@astrophysik.uni-kiel.de}, 
Sebastian Wolf$^{1}$, Alexander V. Krivov$^{2}$, Harald Mutschke$^{2}$,\newauthor and Robert Brunngr\"aber$^{1}$
\\
$^{1}$Kiel University, Institute of Theoretical Physics and Astrophysics, Leibnizstra\ss e 15, 24118 Kiel, Germany\\
$^{2}$Friedrich Schiller University Jena, Astrophysical Institute and University Observatory,  Schillerg\"a\ss chen 2-3, 07745 Jena, Germany
}

\date{Accepted 2017 January 20. Received 2017 January 20; in original form 2016 July 18}

\pubyear{2016}

\begin{document}
\label{firstpage}
\pagerange{\pageref{firstpage}--\pageref{lastpage}}
\maketitle

\begin{abstract}
Recent interferometric surveys of nearby main-sequence stars show a faint but significant near-infrared excess in roughly two dozen systems, i.$\,$e. around $\unit[10]{\%}$ to $\unit[30]{\%}$ of stars surveyed. This excess is attributed to dust located in the immediate vicinity of the star, the origin of which is highly debated. We used previously published interferometric observations to constrain the properties and distribution of this hot dust.
Considering both scattered radiation and thermal reemission, we modelled the observed excess in nine of these systems. We find that grains have to be sufficiently absorbing to be consistent with the observed excess, while dielectric grains with pure silicate compositions fail to reproduce the observations. The dust should be located within $\sim\,\unit[0.01-1]{au}$ from the star depending on its luminosity. Furthermore, we find a significant trend for the disc radius to increase with the stellar luminosity. The dust grains are determined to be below $\unit[0.2-0.5]{\text{\textmu}\textrm{m}}$, but above $\unit[0.02-0.15]{\text{\textmu}\textrm{m}}$ in radius. The dust masses amount to $\unit[(0.2-3.5)\times10^{-9}]{M_\oplus}$. The near-infrared excess is probably dominated by thermal reemission, though a contribution of scattered light up to $\unit[35]{\%}$ cannot be completely excluded. The polarisation degree predicted by our models is always below $\unit[5]{\%}$, and for grains smaller than $\unit[\sim0.2]{\text{\textmu}\textrm{m}}$ even below $\unit[1]{\%}$. We also modelled the observed near-infrared excess of another ten systems with poorer data in the mid-infrared. The basic results for these systems appear qualitatively similar, yet the constraints on the dust location and the grain sizes are weaker.
\end{abstract}




\section{Introduction}
\label{101}

The zodiacal dust cloud in the solar system has been studied intensively by in situ spacecraft measurements, remote observations at visible and infrared wavelengths, and theoretical modelling
(e.$\,$g.~\citealt{Mann2004}).
Dust is present at all heliocentric distances, down to a few solar radii, where it is seen as the F-corona (e.$\,$g.~\citealt{Kimura1998, Hahn2002}). The total zodiacal dust mass has been estimated to $\unit[10^{-9}{-}10^{-7}]{M_\oplus}$ (\citealt{Leinert1996}; \citealt{Fixsen2002}). For comparison, the dust mass of the Kuiper belt amounts to $\unit[(3{-}5)\times 10^{-7}]{M_\oplus}$ (while the total mass in the Kuiper-belt objects is $\unit[\sim0.12]{M_\oplus}$; \citealt{Vitense2012}).
Most of the zodiacal dust is believed to originate from desintegration of short-period comets, with some contribution from asteroids \citep{Nesvorny2010}.

Zodiacal dust around stars other than the Sun, called exozodiacal dust, was first detected around Vega by a deficit of the interferometric visibility compared to that expected from the star alone. Excluding close companions, stellar oblateness or gas emission, such a visibility deficit in the near-infrared (NIR) K waveband (\citealt{Absil2006}) is attributed to thermal reemission of hot dust or scattering of stellar radiation on these grains (e.$\,$g.~\citealt{Akeson2009, Defrere2011, Roberge2012}). So far, an exozodiacal dust environment has been detected through a signature in the NIR around about two dozens of main-sequence stars (\citealt{Absil2009, Absil2013, Ertel2014b}).

The exozodis have been inferred to be by about three order of magnitude brighter than the zodiacal cloud in the solar system. In contrast to the solar system's zodiacal cloud, their origin remains unclear.
Also unclear are the mechanisms that sustain the exozodiacal dust at the observed level. Poynting-Robertson effect, stellar radiation blow-out, and grain collisions are only a small selection of processes which should remove the dust and disperse the disc within a few million years (e.$\,$g.~\citealt{Wyatt2008, Krivov2010, Matthews2014}). At small distances to the star, dispersal mechanisms have particularly short time scales, making the presence of exozodiacal dust at old ages a conundrum (e.$\,$g. \citealt{Rieke2005}).

\begin{figure} 
 \centering
 \includegraphics[trim=1.75cm 1.95cm 0.8cm 2.3cm, clip=true,width=1.0\linewidth]{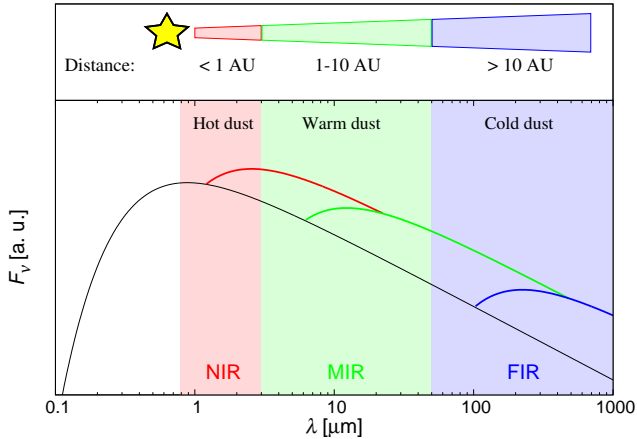}
\vspace*{-0.3cm}
\caption{Illustration of so-called hot, warm, and cold dust traced through their contributions to different parts of the SED. The stellar photosphere (black solid line) corresponds to a solar type star ($T_\star=\unit[5777]{K}$). The height of the excess curves is arbitrary.}
\label{SED_model} 
 \end{figure}

Compared to cold dust located in Kuiper belt analogues at radii larger than $\unit[10]{au}$, causing an excess in the mid- (MIR) and far-infrared (FIR) wavelength range, NIR~observations trace the disc region within or around 1--$\unit[10]{au}$ from the central star (see Fig.~\ref{SED_model}). Thus studies of exozodis offer a way to better understand the inner regions of extrasolar planetary systems. Besides, 
the possible presence of small grains in exozodiacal clouds is a potential problem for the detection of 
terrestrial planets in the habitable zone of these systems (e.$\,$g.~\citealt{Agol2007, Beckwith2008}). This provides another motivation to study the exozodiacal dust and its properties.

Previously published constraints favour discs consisting of small particles in a narrow ring around the star (e.$\,$g.~\citealt{Rieke2016}). The dust has to be located beyond the sublimation radius of the grains, which typically amounts to a few solar radii (\citealt{Mann2004}). For Vega, Fomalhaut, and $\beta$~Leo, small refractory particles ($\sim\unit[10-500]{nm}$) at distances of $\unit[\sim0.1-0.3]{au}$ in narrow rings are required to account for the hot excess (\citealt{Absil2006, Akeson2009, Defrere2011, Lebreton2013}).
\citet{Rieke2016} obtained similar results, constraining the maximum grain size to $\sim\unit[200]{nm}$.

Due to the proximity to the star, spatially resolved direct imaging observations of the exozodiacal dust hosting environments are missing. Although \cite{Rieke2016} found a weakly but significantly bluer photometric colour for these systems compared to stars without exozodiacal dust, the detection of these systems was only possible so far through interferometric observations. The observing strategy is to measure the visibility using a suitable set of baseline configurations. A significant deficit of the interferometric visibility at low spatial frequencies indicates the presence of dust material spatially resolved and beyond their sublimation distance, enabling direct determination of the flux ratio between  disc and central star (\citealt{diFolco2004}). Two interferometric surveys using CHARA/FLUOR (\citealt{Absil2013}) and VLTI/PIONIER (\citealt{Ertel2014b}) resulted in detections of exozodiacal dust around main-sequence stars with an incidence rate of $\unit[10]{\%}$ to $\unit[30]{\%}$.

Based on the observational data available, we intend to put stronger constraints on the exozodiacal dust distribution than those that were derived previously. For this purpose, in this paper we model selected NIR to FIR observations of systems for which NIR long-baseline interferometric observations indicate the presence of hot dust close to the star.
Our sample of systems and the methods we use to analyse their observations are described in Sect.~\ref{201}. The results are presented in Section~\ref{301}.
A discussion
of possible mechanisms that may create and sustain the exozodiacal dust
is given in Section~\ref{401}. Section~\ref{501} summarises our findings.

\section{Analysis}
\label{201}
In this section we present both our selection of systems and the methods we use to constrain the parameters of the spatial distribution of the dust and the sizes of the dust grains. The survey of stars with hot exozodiacal dust is introduced in Section~\ref{sec_survey} and the disc and dust model are described in Section~\ref{sec_model}. The procedure and the observational data applied to constrain the model parameters are presented in Section~\ref{sec_proceed} and \ref{sec_data}, and the calculation of the sublimation radius is described in Section~\ref{sec_sublim}. A discussion of the impact of the limited spatial resolution of the NIR and MIR~observations is given in Section~\ref{sec_innerwork}.

\subsection{Survey of stars with NIR~excess}
\label{sec_survey}
In Table~\ref{data_systems} those systems of the surveys of \cite{Absil2013} in K~band and \cite{Ertel2014b} in H~band are compiled which show a significant NIR~excess that is indicative of hot exozodiacal dust. In addition, HD~216956 (Fomalhaut) is included, which was observed with VLTI/VINCI in K~band (\citealt{Absil2009}). 
Stars for which a detected NIR~excess can be attributed to a different mechanism or a companion (\citealt{Mawet2011, Absil2013, Bonsor2013a, Marion2014}) have been excluded from this study. Besides \text{HD~7788} ($\kappa$~Tuc), for which a temporal variability of the excess was detected (\citealt{Ertel2016}), the list of known systems with hot exozodiacal dust is complete. \cite{Ertel2016} also found a tentative indication for a temporal variability of the excess of HD~210302 ($\tau$~PsA). However, since this variability is not significant, the system is considered in our study. All of the listed stars are at or close to the main sequence, with spectral classes ranging from A0 to G8 and luminosities from $L_\star\sim\unit[0.4]{L_\odot}$ to $\unit[\sim40]{L_\odot}$. 

To derive the flux of the NIR~excess from the observed visibilities, uniformly bright emission was adopted in the whole field of view (\citealt{diFolco2004}). We divide the selected systems in two groups: Group~I comprises all systems with interferometric flux measurements at $\lambda=\unit[8.5]{\text{\textmu}\textrm{m}}$ which will help to find strong constraints for the disc model. Group~II contains all systems without such observations at $\lambda=\unit[8.5]{\text{\textmu}\textrm{m}}$. As we will see in Section~\ref{survey_se}, the dust parameters in these systems cannot be constrained as tightly as for the Group~I-targets.

\begin{table*} 
\caption{Parameters of targets with NIR excess (\citealt{Absil2009}; \citealt{Absil2013}; \citealt{Ertel2014b}) which is attributed to circumstellar dust. Group~I/Group~II comprises all systems with/without interferometric observations at $\lambda=\unit[8.5]{\text{\textmu}\textrm{m}}$ (\citealt{Mennesson2014}). \newline \textbf{Notes to the IR-excesses:} ``H'' excess in the H~band; ``K'' excess in the K~band; ``(K)'' observed in K band, but without information on the excess (\citealt{Defrere2012}); ``/'' no interferometric observation; ``-'' no significant excess detected; ``(+)'' excess uncertain, discussed differently in literature, SED makes no clear statement; ``+'' weak significant excess; ``++'' strong significant excess}
\label{data_systems}
 \begin{tabular}{r r c| c c c c c| c c c c c} \hline\hline
HD\hspace*{0.2cm}& HIP\hspace*{0.2cm}& Alter.& $d$  & $T_\star$& $L_\star$ &Spectral   &Age&NIR&$\unit[8.5]{\text{\textmu}\textrm{m}}$&$\unit[24]{\text{\textmu}\textrm{m}}$&$\unit[70]{\text{\textmu}\textrm{m}}$&Ref.\\
number& number &name& [pc]                             &$\!$[K]              &[$\text{L}_\odot$]& class          &[Gyr]&excess&excess&excess&excess&excess\\\hline
\multicolumn{12}{l}{\hspace*{0.7cm}\underline{Group~I}}\\
10700&8102&$\tau$~Cet	&$\phantom{1}3.7$   	&$\phantom{1}$5290&$\phantom{1}0.46$&G8$\,$V        	  	&$10\phantom{.00}$&K&-&-&+&(a), (b), (c), (d)\\
22484&16852&10~Tau		&$13.7$          	&$\phantom{1}$5998&$\phantom{1}3.06$&F9$\,$IV-V    	&$6.7$&K&-&-&+&(d), (e), (f)\\
56537&35350&$\lambda$~Gem	&$28.9$       		&$\phantom{1}$7932&$27.4\phantom{0}$&A3$\,$V      	&$0.5$  &K&-&-&-&(d), (g), (h)\\
102647&57632&$\beta$~Leo	&$11.1$			&$\phantom{1}$8604&$13.25$	    &A3$\,$V        	&$0.1$  &K&+&+&++&(d), (i)\\
172167&91262&$\alpha$~Lyr	&$\phantom{1}7.8$       &$\phantom{1}$9620&$37\phantom{.70}$&A0$\,$V	 	&$0.7$  &H, K&-&-&++&(d), (i), (j), (k)\\
177724&93747&$\zeta$~Aql	&$25.5$                 &$\phantom{1}$9078&$36.56$	    &A0$\,$IV-V	        &$0.8$  &K&-&-&-&(d), (l), (m)\\
187642&97649&$\alpha$~Aql	&$\phantom{1}5.1$       &$\phantom{1}$7680&$10.2\phantom{0}$&A7$\,$IV-V		&$1.3$  &K&-&-&-&(c), (d), (g), (n)\\
203280&105199&$\alpha$~Cep	&$15.0$                 &$\phantom{1}$7700&$19.97$          &A7$\,$IV-V		&$0.8$  &K&-&-&-&(d), (l)\\ 
216956&113368&$\alpha$~PsA	&$\phantom{1}7.7$                  &$\phantom{1}$8590&$16.6\phantom{0}$          &A3$\,$V		&$0.4$  &K&-&+&++&(d), (o), (p), (q)\\ 
\multicolumn{12}{l}{\hspace*{0.7cm}\underline{Group~II}}\\
 2262&2072&$\kappa$~Phe	&$23.5$                 &$\phantom{1}$9506&16$\phantom{.38}$&A5$\,$IV            &$0.7$&H&/&+&+&(i)\\
14412&10798&			&$12.7$                 &$\phantom{1}$5371&$\phantom{1}0.38$&G8$\,$V             &$3.7$ &H&/&-&-&(f)\\   
20794&15510&e Eri		&$\phantom{1}6.1$       &$\phantom{1}$5401&$\phantom{1}0.66$&G8$\,$V		 &$8.1$ &H&/&(+)&(+)&(r), (s)\\
28355&20901&b~Tau		&$49.2$                 &$\phantom{1}$7852&16$\phantom{.38}$&A7$\,$V		 &$0.7$ &H&/&+&++&(i)\\
39060&27321&$\beta$~Pic	&$19.3$                 &$\phantom{1}$8203&13$\phantom{.38}$&A6$\,$V		 &$\phantom{2}0.02$ &H, (K)&/&++&++&(i)\\
104731&58803&			&$24.2$                 &$\phantom{1}$6651&$\phantom{1}$4$\phantom{.38}$&F5$\,$V  &$1.6$&H&/&-&-&(f)\\
108767&60965&$\delta$~Crv	&$26.9$                 &            10209&40$\phantom{.38}$&A0$\,$IV		  &$0.3$&H&/&-&-&(i)\\
131156&72659&$\xi$~Boo	&$\phantom{1}6.5$       &$\phantom{1}$5483&$\phantom{1}0.6\phantom{0}$&G7$\,$V	  	&$0.3$&K&/&-&-&(g)\\
173667&92043&110~Her		&$19.1$                 &$\phantom{1}$6296&$\phantom{1}6.01$&F5.5$\,$IV-V	  &$3.4$&K&/&-&(+)&(e), (s)\\
210302&109422&$\tau$~PsA	&$18.7$                 &$\phantom{1}$6339&$\phantom{1}2.5\phantom{0}$&F6$\,$V    &$2.2$&H&/&-&-&(e), (f)\\\hline 
 \end{tabular}
\newline
\raggedright
\textbf{References:} (a)~\cite{Greaves2004}; (b)~\cite{Chen2006}; (c)~\cite{Habing2001}; (d)~\cite{Mennesson2014}; (e)~\cite{Kospal2009}; (f)~\cite{Trilling2008}; (g)~\cite{Gaspar2013}; (h)~\cite{Chen2014}; (i)~\cite{Su2006}; (j)~\cite{Gillett1986}; (k)~\cite{Defrere2011}; (l)~\cite{Chen2005}; (m)~\cite{Plavchan2009}; (n)~\cite{Rieke2005};  (o)~\cite{Stapelfeldt2004}; (p)~\cite{Su2013};  (q)~\cite{Acke2012}; (r)~\cite{Wyatt2012}; (s)~\cite{Beichman2006}. The distances are derived from parallax measurements (\citealt{vanLeeuwen2007}). Stellar temperatures and luminosities are adopted from (c), \cite{vanBelle2001}, \cite{vanBelle2006}, \cite{Wyatt2007}, \cite{Mueller2010}, \cite{Pepe2011}, \cite{Zorec2012}, \cite{Boyajian2013}, and \cite{Pace2013}, spectral classes and stellar ages are taken from \cite{Mamajek2012}, \cite{Vican2012}, \cite{Absil2013}, \cite{Ertel2014b}, and \cite{Mamajek2014}.
\end{table*}

\subsection{Model description}
\label{sec_model}
The aim of the study is to fit the NIR~fluxes of the hot exozodiacal dust systems.
Owing to a paucity of observational data, a simple disc model is adopted to keep the number of free parameters small. 

{\it Disc properties.} The disc model represents a thin ring with an inner radius $R$ and outer radius $R_\text{out}=1.5\,R$. The number density decreases with $n(r)\propto r^{-1}$. The dust mass $M_{\mathrm{dust}}$ is chosen as a scaling factor to match the NIR~flux for each individual system. Taking into account scattered radiation, the inclination $i$ of the disc has to be considered. We examine three cases: a face-on ($i=0^\circ$) and an edge-on disc ($i=90^\circ$) with an half opening angle of $5^\circ$, and a spherical shell around the central star. The latter case is motivated by the scenario of dust expelled from exo-Oort cloud comets, or by stellar magnetic fields which compel charged dust grains on orbits perpendicular to the disc plane.

{\it Dust properties.} Each dust ring is composed of compact and spherical dust grains with a single grain radius $a$. The grains are assumed to consist of pure graphite ($\rho=2.24\,$g\,cm$^{-3}$; \citealt{WeingartnerDraine2001}), using the 1/3$-$2/3 approximation (\citealt{DraineMalhotra1993}). 

The optical properties of the grains are calculated with the software tool \texttt{miex} which is based on the theory of Mie-scattering (\citealt{Mie1908}; \citealt{WolfVoshchinnikov04}). Single scattering and reemission simulations are performed and the SEDs are calculated using an enhanced version of the tool \texttt{debris} (\citealt{Ertel2011}). All free parameters and the range within which they have been varied are presented in Table~\ref{free_parameter}.

 \begin{table}
 \caption{Parameter space for modelling the systems from Tab.~\ref{data_systems}.}
 \centering
 \label{free_parameter}
 \begin{tabular}{l c l c c c} 
\hline\hline
\multicolumn{3}{c}{Parameter}   &Range&\hspace*{-0.5cm}Number&\hspace*{-0.3cm}Grid\\\hline
Disc ring radius$\hspace*{-0.3cm}$&$R\hspace*{-0.3cm}$& $[$au$]\hspace*{-0.3cm}$ &$0.01\phantom{0}-10\phantom{.}$ &\hspace*{-0.5cm}100&\hspace*{-0.3cm}log.\\
Grain size$\hspace*{-0.3cm}$&$a\hspace*{-0.3cm}$& $[\text{\textmu}\textrm{m}]\hspace*{-0.3cm}$&$0.001-10/100$       &\hspace*{-0.5cm}100&\hspace*{-0.3cm}log.\\ 
Inclination $\hspace*{-0.3cm}$&$i\hspace*{-0.3cm}$&                  & $0^\circ$, $90^\circ$, sphere&\hspace*{-0.5cm}3&\hspace*{-0.3cm}discr.\\
\hline
 \end{tabular}
\end{table}


\subsection{Procedure}
\label{sec_proceed}
\begin{figure}
 \includegraphics[trim=1.7cm 1.95cm 1.6cm 2.0cm, clip=true,page=1,width=1.0\linewidth]{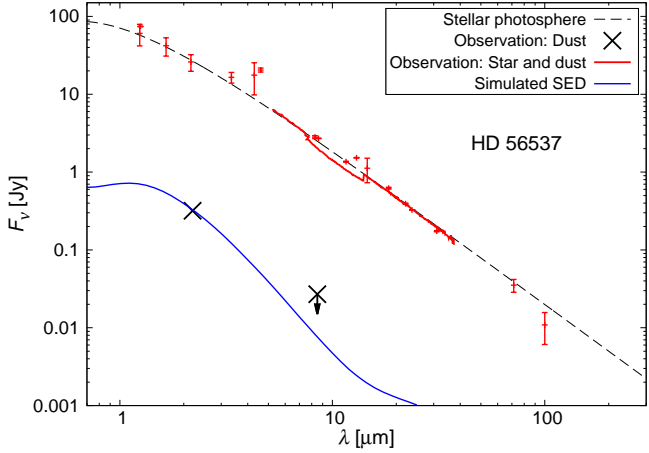}
\caption{SED of HD~56537. The dashed line represents the stellar photosphere ($T_\star=\unit[7932]{K}$, $L_\star=\unit[27.4]{L_\odot}$), the red points show the observed fluxes (star plus dust, including Spitzer/IRS spectrum; \citeauthor{VIZIER}-catalogue) and the black points mark the excess of the dust (without stellar flux contribution). The flux at \mbox{$\lambda=\unit[8.5]{\text{\textmu}\textrm{m}}$} is an upper limit. A simulated SED which reproduces the NIR flux without overestimating the observed fluxes in the MIR and FIR is shown as the blue solid line. }
\label{SED_HD2262}
\end{figure}

To illustrate the data modelling procedure, the SED of the NIR excess harbouring system HD~56537 is shown in Figure~\ref{SED_HD2262}. Obviously, many appropriate parameter settings exist which would allow one to reproduce the NIR~excess. Fortunately, there exist further observations which potentially help to constrain the possible ranges for the grain size $a$ and the disc ring radius $R$. In particular, the observed MIR and FIR~fluxes provide strict upper limits on the total mass (for given dust properties and radial density distribution). Furthermore, an underestimation of the observed MIR~ and FIR~fluxes would still be in agreement with the observations
because of the potential presence of additional dust distributions located at larger distances to the star, similar to the Kuiper~belt in the solar system.

The ratio of the simulated flux in the NIR to the simulated flux in the MIR and FIR is determined for each parameter setting:
\begin{equation}
 S^{\text{simu}}_{\lambda_\text{MIR/FIR}}=\frac{F^{\text{simu}}_\nu(\lambda_\text{NIR})}{F^{\text{simu}}_\nu(\lambda_\text{MIR/FIR})}.\label{coloreq}
\end{equation}
The simulated colour $S^{\text{simu}}_{\lambda_\text{MIR/FIR}}$ has to be compared to the observed colour,
\begin{equation}
S^{\text{obs}}_{\lambda_\text{MIR/FIR}}=\frac{F^{\text{obs}}_\nu(\lambda_\text{NIR})}{F^{\text{obs}}_\nu(\lambda_\text{MIR/FIR})},\label{Sobs}
\end{equation}
and must fulfill the condition 
\begin{equation}
S^{\text{simu}}_{\lambda_\text{MIR/FIR}}\,\ge\, S^{\text{obs}}_{\lambda_\text{MIR/FIR}}\label{help}
\end{equation}
in order to reproduce the NIR~excess. In this study we assume that the hot dust emission does not vary with time (see Sect.~\ref{sec_time_variab} for a discussion of the time variability of the NIR~excess).

\subsection{Observational data}
\label{sec_data}

\subsubsection{NIR~data}
Except for HD~216956 (Fomalhaut), the NIR~fluxes of the dust around all Group~I-targets as well as HD~131156 and HD~173667 of Group~II are determined by \text{FLUOR/CHARA} (\citealt{Absil2013}). The data for HD~216956 are taken with \text{VLTI/VINCI} (\citealt{Absil2009}), while HD~14412, HD~104731 and HD~108767 of Group~II are observed with \text{VLTI/PIONIER} (\citealt{Ertel2014b}). For the remaining targets of Group~II, the weighted mean flux ratios of \cite{Ertel2014b, Ertel2016} are used (\text{PIONIER}).

\subsubsection{MIR and FIR~data}
The colours (Eq.~\ref{coloreq}) are analysed for the wavelengths \mbox{$\lambda=\unit[24]{\text{\textmu}\textrm{m}}$} and $\unit[70]{\text{\textmu}\textrm{m}}$ for all Group~I and II-targets. Table~\ref{data_systems} indicates which of the systems show a significant excess in the MIR or FIR. Several systems have no or only minor excess at these wavelengths. To cope with the problem that a determination of the flux emitted by the dust is affected mainly by the uncertainty of the stellar photospheric flux, we use the measured flux which comprises the contribution of both dust and stellar photosphere at $\lambda_\text{MIR/FIR}=\unit[24]{\text{\textmu}\textrm{m}}$ and $\unit[70]{\text{\textmu}\textrm{m}}$. Although this condition is weaker than Eq.~(\ref{help}), this helps to circumvent the uncertainty of the photospheric flux.

For HD~10700, we use IRAS-fluxes at $\unit[60]{\text{\textmu}\textrm{m}}$ and  $\unit[100]{\text{\textmu}\textrm{m}}$ (\citealt{Greaves2004}) to interpolate the excess flux at \mbox{$\lambda=\unit[70]{\text{\textmu}\textrm{m}}$} since there exists no observation at this wavelength. For HD~131156 and HD~187642, both fluxes at $\unit[24]{\text{\textmu}\textrm{m}}$ and  $\unit[70]{\text{\textmu}\textrm{m}}$ are estimated from the stellar photosphere and interpolation of observations of IRAS (\citealt{Habing2001}), Akari, WISE, and Herschel/PACS at $\unit[100]{\text{\textmu}\textrm{m}}$ (\citealt{Gaspar2013}). 

In addition, we use MIR~data from Keck for the Group~I-targets (\citealt{Mennesson2014}). For each of these systems we calculate the flux emitted from the dust at $\lambda=\unit[8.5]{\text{\textmu}\textrm{m}}$  using $F_\text{dust}\sim2.5\,F_\star\,E_\text{8-9}$ (B. Mennesson, private communication), where $E_\text{8-9}$ is the measured excess leak given in their Table~2. The factor $2.5$ corresponds to the average transmission value over the region extending from the inner and outer working angle. Only one system (HD~102647) shows a significant excess. For all the other systems we use the significance limit as an upper limit for the flux at $\lambda=\unit[8.5]{\text{\textmu}\textrm{m}}$.

\subsection{Sublimation radius and habitable zone}
\label{sec_sublim}
Dust sublimation in the vicinity of the star sets the minimum distance at which the dust material can be located. Subsequently to the investigation of the parameter space presented in Section~\ref{sec_model}, we determine the sublimation radii as a function of grain size $a$ in thermal equilibrium. The sublimation radius $R_\text{sub}$ of a particle with a sublimation temperature $T_\text{sub}$ amounts to
\begin{equation}
R_\text{sub}=\frac{R_\star}{2}\sqrt{\frac{\int_0^\infty B_{\lambda} \left(T_\star\right) Q_{\mathrm{abs}}\left(a,\lambda\right)  \,\mathrm d\lambda }{\int_0^\infty B_{\lambda}\left(T_{\mathrm{sub}}\right) Q_{\mathrm{abs}}\left(a,\lambda\right)  \,\mathrm d\lambda }}\, ,\label{AbstanddT}
\end{equation}
(\citealt{Backman1993}), where $B_\lambda$ is the Planck function and $Q_\text{abs}$ is the dimensionless absorption efficiency. We assume a sublimation temperature of $T_\text{sub}=\unit[2000]{K}$ (e.$\,$g.~\citealt{Lamoreaux1987}) for graphite and determine the sublimation radius for each system. It should be noted that the modelling results presented in Section~\ref{301} strongly depend on the exact choice of the sublimation temperature $T_\text{sub}$.

For an approximate estimation of the borders of the habitable zone we assume a temperature profile
\begin{equation}
T_\text{HZ}=(\unit[278.3]{K})\left(\frac{L_\star}{L_\odot}\right)^{0.25}\left(\frac{\unit[1]{au}}{R_\text{HZ}}\right)^{0.5}, \label{HabiZone}
\end{equation}
where $T_\text{HZ}$ is the temperature at the radial distance  $R_\text{HZ}$ of the habitable zone. The temperature range is 210 to $\unit[320]{K}$, equivalent to the habitable zone location in the solar system (e.$\,$g.~\citealt{Kopparapu2013}).
\subsection{Spatially resolved dust emission and sensitivity dependency over the field of view}
\label{sec_innerwork}
The circumstellar emission in the NIR is spatially resolved. However, dust within an inner working angle $\theta$ does not contribute to the NIR~excess, which has to be considered in the modelling.

The angular resolution of an interferometer is given by
\begin{align}
\theta=\lambda/(x\,B),\label{resolution_inter}
\end{align}
where $B$ is the projected baseline and $x\in\left\lbrace1,\,2,\,4,\,...\right\rbrace$. Since the correct value for $x$ is contentious,  we choose a different approach. We assume a uniformly bright emission ring (face-on)\footnote{Similarly, a uniformly bright emission is also adopted in the whole field of view by \cite{Absil2009}, \cite{Absil2013} and \cite{Ertel2014b} to derive the NIR~flux of the dust from the observations.} with an inner radius $R$ and outer radius $R_\text{out}=1.5\,R$ and adopt the measured NIR~flux ratio $f_\text{observation}$ of the spatially resolved dust and the central star. By varying the inner radius, we calculate the visibility function $V_{\text{(}\star\text{ + disc)}}$ of the star in combination with the disc for this simplified system. Subsequently, the ratio $f'$ of the simulated observational date is derived via $f'=(V_\star-V_{\text{(}\star\text{ + disc)}})/V_\star$,  where $V_\star$ denotes the stellar visibility alone (\citealt{diFolco2004}).

\begin{figure}
 \includegraphics[trim=1.7cm 0.65cm 1.6cm 2.0cm, clip=true,page=1,width=1.0\linewidth]{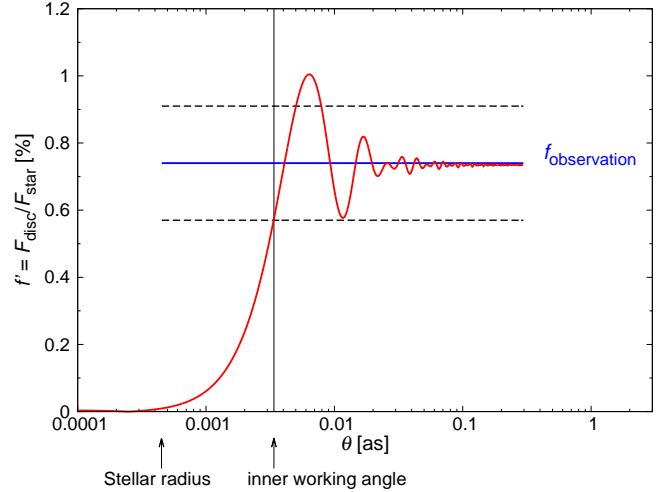}
\caption{Ratio $f'=F_\text{disc}/F_\star$ for the system HD~56537, derived from the simulated visibility function of a uniformly bright emission ring (face-on) for different inner radii $R$ (see Sect.~\ref{sec_innerwork} for details). The inner working angle of the simulated NIR~observation is defined as the smallest value $\theta$ for which the ratio $f'$ is still within the interval $f_\text{observation}\pm\sigma_f$, where $\sigma_f$ is the uncertainty of the real NIR~observation.}
\label{Derive_f}
\end{figure}

For large inner radii (\textit{right} in Figure~\ref{Derive_f}), the entire disc is spatially resolved, and the derived ratio $f'$ equals the given ratio $f_\text{observation}$. Decreasing the inner radius, the emission gets less resolved, resulting in a modulation of $f'$ and an increase of the difference between $f'$ and $f_\text{observation}$. For small inner radii  (Fig.~\ref{Derive_f}, \textit{left}), the entire disc is unresolved and $f'$ converges to 0. We define the inner working angle of the simulated NIR~observation as the smallest value $\theta$ for which the derived ratio $f'$ is still within the interval $f_\text{observation} \pm\sigma_f$, where $\sigma_f$ is the uncertainty of the real NIR~observation. This is the most conservative approach for the adoption of an inner working angle, since we want to avoid assuming a too large inner working angle and thus constraining the dust distribution in a wrong way. For the \text{FLUOR} observation of the system HD~56537 in Figure~\ref{Derive_f}, a value of $\theta=\unit[0.0034]{as}$ is derived. Using Eq.~\ref{resolution_inter}, this is equivalent to $x=3.97$. For comparison, \cite{Absil2013} assumed a value of $4$ when examining the resolution of the sublimation radius. In our study, $x$ is  between $3.4$ and $4.1$ for all observations with \text{FLUOR}, \text{PIONIER} and \text{VINCI} of the systems in Table~\ref{data_systems}. The small variability in $x$ is the result of the uncertainty $\sigma_f$ of the individual observations.

Additionally to the interferometric data in the NIR, MIR~nulling data obtained with the Keck Interferometer are used for the Group~I-targets. The inner working angle for these observations is $\unit[6]{mas}$ (\citealt{Mennesson2014}). The signal for sources at projected separations larger than $\unit[200]{mas}$ is strongly attenuated or even completely missed, setting an outer working angle of the instrument.

After the determination of the inner working angle for each system and NIR and MIR~observation, it is used as a strong limiting condition for the simulations: radiation which is scattered or reemitted within the inner working angle (or outside the outer working angle) is completely neglected. While in the face-on case only discs with small inner radii $R$ are affected, a significant fraction of radiation is never resolved by the case of edge-on discs and spherical dust shells. In particular, a strong forward-scattering peak is invisible if scattered light is considered.

In addition to the inner working angle, the sensitivity of an optical, single mode fiber interferometer such as \text{FLUOR}, \text{PIONIER}, or \text{VINCI} is not uniform over the field of view. Instead, it can be approximated as a Gaussian function with a FWHM of $\unit[0.8]{as}$, $\unit[0.4]{as}$, and $\unit[1.6]{as}$, respectively (\citealt{Absil2013, Ertel2014b, Absil2009}). This disparity is taken into account in our NIR~simulations, so that emission further away from the star is weaker than emission closer in.

\section{Results}
\label{301}
 \begin{figure*}
\vspace*{-0.2cm} \includegraphics[trim=2.1cm 14.65cm 2.4cm 4.550cm, clip=true,width=1.01\linewidth, page=1]{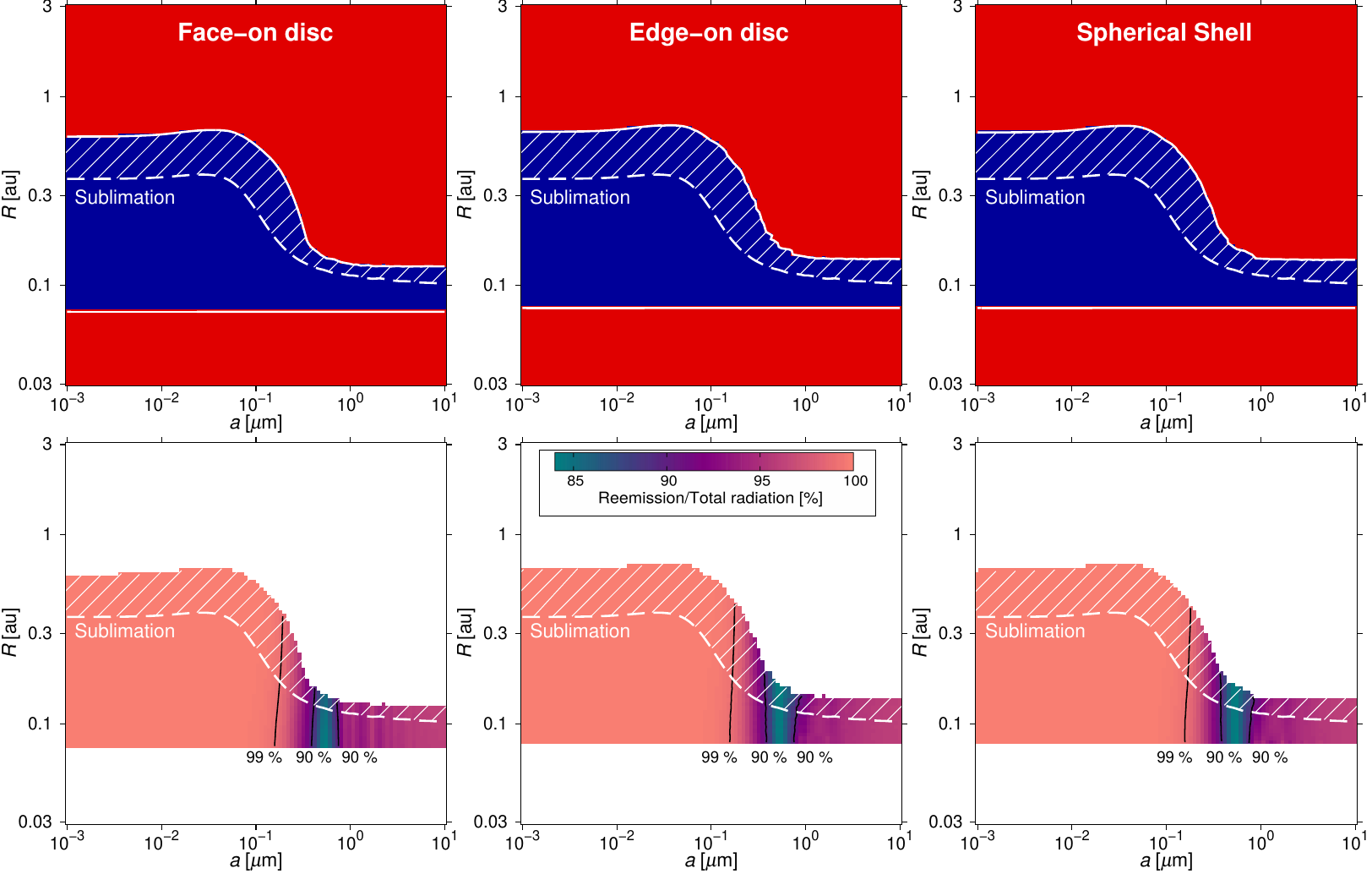}\vspace*{-0.2cm}
\caption{Explored parameter space for the system HD~56537. The dust material is pure graphite. The disc inclination is $i=0^\circ$ (\textit{left column}) and $90^\circ$ (\textit{middle column}), and the results for the case of a spherical dust distribution are shown in the \textit{right column}. In the \textit{first row} the calculations considering the observations at $\lambda=\unit[2.2]{\text{\textmu}\textrm{m}}$, $\unit[8.5]{\text{\textmu}\textrm{m}}$, $\unit[24]{\text{\textmu}\textrm{m}}$ and $\unit[70]{\text{\textmu}\textrm{m}}$ are shown. The blue regions in the panels correspond to parameter settings which can reproduce the observational data, while the parameter settings of the red regions fail. The sublimation radius (white dashed line) and the upper blue-red border comprise the region of suitable parameter settings (dashed area). In the \textit{second row} the ratio of the reemission to the total radiation is illustrated. The colour scale is linear, ranging from $\unit[84]{\%}$ (turquoise) up to $\unit[100]{\%}$ (apricot). Black contour lines show the $\unit[90]{\%}$ and $\unit[99]{\%}$ level.}
\label{Para1a} 
 \end{figure*}
In this section the results of the disc modelling process are presented. For each parameter setting in Table~\ref{free_parameter} the colour $S^{\text{simu}}_{\lambda_\text{MIR/FIR}}$ is calculated and compared to $S^\text{obs}_{\lambda_\text{MIR/FIR}}$. At first we demonstrate the key results of the modelling for the exozodiacal dust system HD~56537 as a representative of Group~I in detail (Sect.~\ref{represent}). Subsequently, we discuss the entire samples of Group~I and II (Sect.~\ref{survey_sd} - \ref{survey_se}). In Section \ref{sec_uncertainty} - \ref{sec_disp} the uncertainty and the time variability of the NIR~excess as well as constraints by spectrally dispersed data are discussed.

\subsection{HD~56537}
\label{represent}
HD~56537 ($\lambda$~Gem) is a $\unit[5\times10^8]{year}$ old system (\citealt{Vican2012}) with a significant K band excess ($\unit[0.74]{\%}\pm\unit[0.17]{\%}$ of the stellar flux), but with no significant excess at $\lambda=\unit[8.5]{\text{\textmu}\textrm{m}}$, $\unit[24]{\text{\textmu}\textrm{m}}$ or $\unit[70]{\text{\textmu}\textrm{m}}$. This suggests that the NIR~excess stems solely from the hot exozodiacal dust, with no contribution from other potential disc components. The results of the modelling are shown in Figure~\ref{Para1a}. Each panel represents a selected range of the parameter space consisting of the grain size $a$ and the dust ring radius $R$. In the following, we investigate potential constraints which can be derived for these quantities and determine the contribution of scattered and polarised radiation of the NIR flux.
\subsubsection{Constraints on grain size and dust location}
The results for different  disc inclinations ($i=0^\circ$ or $90^\circ$) and for the spherical dust shell  are presented in Figure~\ref{Para1a}. Small differences  between these three optically thin cases  exist which are due to the dependence of the scattered NIR and MIR radiation on the light scattering geometry. In addition, the fraction of radiation from dust close to the star (which does not contribute to the observed flux) as well as the fraction of radiation from dust located further away (which does not change the observed flux either, due to the sensitivity of the optical single mode fiber) depend on the dust distribution geometry.
The results for the spherical shell can be interpreted qualitatively as intermediate between the results for the face-on and the edge-on disc.

In the \textit{upper row} of Figure~\ref{Para1a} the calculations considering the observations at the wavelengths $\lambda=\unit[8.5]{\text{\textmu}\textrm{m}}$, $\unit[24]{\text{\textmu}\textrm{m}}$ and $\unit[70]{\text{\textmu}\textrm{m}}$ are shown. In the blue-coloured regions the corresponding parameter settings reproduce the NIR~excess without overestimating the fluxes at these three wavelengths, while in the red-coloured regions the simulated fluxes of at least one of them is too large. Models with inner radii $R\lesssim\unit[0.08]{au}$ fail because of the inner working angle of the simulated NIR~observations. On the other hand, models with inner radii $\unit[0.08]{au}\lesssim R\lesssim\unit[0.14]{au}$ reproduce the observations because of the larger inner working angle of the simulated MIR~observations. Since the sublimation radius is spatially resolved by the NIR~observations for all grain sizes, it depicts the minimum limit for the inner radius $R$ ($\unit[0.1]{au}$). Furthermore, sublimation radius and the upper blue-red border comprise the region of the suitable parameter settings (dashed area). This area in the parameter space is evaluated for each system and the corresponding parameters for minimum and maximum dust ring radius, grain size, and mass are derived. For HD~56537, the maximum disc ring radius is $R_\text{max}=\unit[0.65]{au}$ ($\unit[0.68]{au}$; $\unit[0.68]{au}$) for the face-on disc (edge-on disc; sphere). The total dust mass required to reproduce the NIR~flux is between $\unit[0.5\times10^{-9}]{M_\oplus}$ ($\unit[0.9\times10^{-9}]{M_\oplus}$; $\unit[0.6\times10^{-9}]{M_\oplus}$) and $\unit[22\times10^{-9}]{M_\oplus}$ ($\unit[52\times10^{-9}]{M_\oplus}$; $\unit[33\times10^{-9}]{M_\oplus}$). The grain sizes for this system are not limited by our study, and both nanometer-sized grains and those larger than $\sim\unit[10]{\text{\textmu}\textrm{m}}$ can reproduce the observed excesses. The reason for the possible presence of the large particles is that the sublimation radius of this system is not spatially resolved by the MIR observation at $\lambda=\unit[8.5]{\text{\textmu}\textrm{m}}$. A pile-up of large grains near the sublimation radius would cause a detectable excess in the NIR, which could not be observed by the MIR~observations. In contrast, the theoretical approach of an infinitely small resolution of both the NIR or MIR~observations would rule out large grains in this system due to small observed fluxes at $\lambda=\unit[8.5]{\text{\textmu}\textrm{m}}$ compared to the NIR.

In general, a higher sublimation temperature would decrease the minimum radius and minimum dust mass and increase the maximum grain size.
 \begin{figure} 
\includegraphics[trim=3.55cm 1.95cm 2.75cm 2.05cm, clip=true,width=1.0\linewidth, page=3]{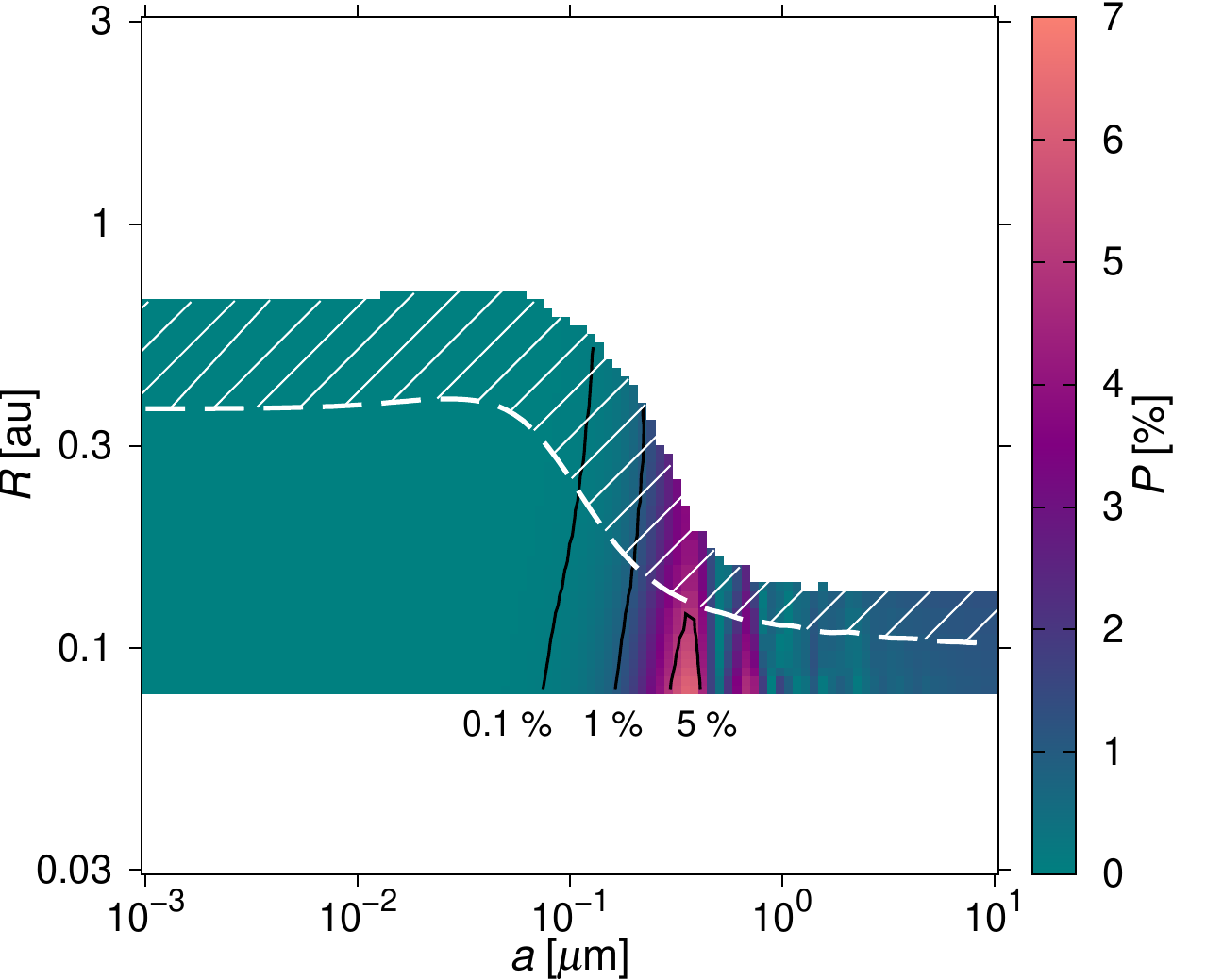} 
\caption{Polarisation degree $P$ of the total radiation for the system HD~56537. The dashed white line indicates the sublimation radius and the region of suitable parameter settings is highlighted as dashed area (see Fig.~\ref{Para1a}). The black contour lines show the $\unit[0.1]{\%}$, $\unit[1]{\%}$  and $\unit[5]{\%}$ level. The dust material is graphite and the inclination is $90^\circ$.}
\label{Result_Polarisation_Silikat}
 \end{figure}

\subsubsection{Scattered and polarised radiation}
The contribution of scattered radiation to the NIR flux is still an open question, although previously published studies tended to favour a small contribution (e.$\,$g. \citealt{vanLieshout2014, Rieke2016}). The ratio of the thermal reemission radiation to the total radiation is illustrated in the \textit{second row} of Figure~\ref{Para1a}. We find that for suitable parameter settings thermal reemission is the dominating source. The lowest ratio of the thermal reemission to the total radiation for all three disc geometries amounts to $\unit[84]{\%}$. For grain radii below $\unit[0.2]{\text{\textmu}\textrm{m}}$ the ratio is larger than $\unit[99]{\%}$. Consequently, the contribution of scattered radiation and thus the influence of the disc inclination are negligible for these small grains. Only for grains with radii between  $\unit[0.2]{\text{\textmu}\textrm{m}}$ and $\unit[\sim 1]{\text{\textmu}\textrm{m}}$, the fraction of scattered radiation is larger.

Moreover, the polarisation degree $P$ of the total radiation in the NIR is determined (Fig.~\ref{Result_Polarisation_Silikat}). According to the low contribution of scattered radiation for the suitable parameter settings, the polarisation degree is below $\unit[5]{\%}$ and even below $\unit[1]{\%}$ for grains smaller than $\unit[0.24]{\text{\textmu}\textrm{m}}$. Only for grain radii between $\unit[0.3]{\text{\textmu}\textrm{m}}$ and $\unit[\sim 0.5]{\text{\textmu}\textrm{m}}$, the polarisation is larger than $\unit[2]{\%}$. Because of symmetry constraints, the integrated polarisation equals zero in the case of the face-on disc and the spherical shell. The results are consistent with polarisation measurements of hot dust stars with the instrument HIPPI at the Anglo-Australian Telescope in the optical wavelength range, which have shown that the detected polarised emission does not exceed $\unit[1]{\%}$ (\citealt{Marshall2016}).

\subsubsection{Different dust materials}
 \begin{figure}
 \includegraphics[trim=1.7cm 1.95cm 1.6cm 2.0cm, clip=true,page=1,width=1.0\linewidth]{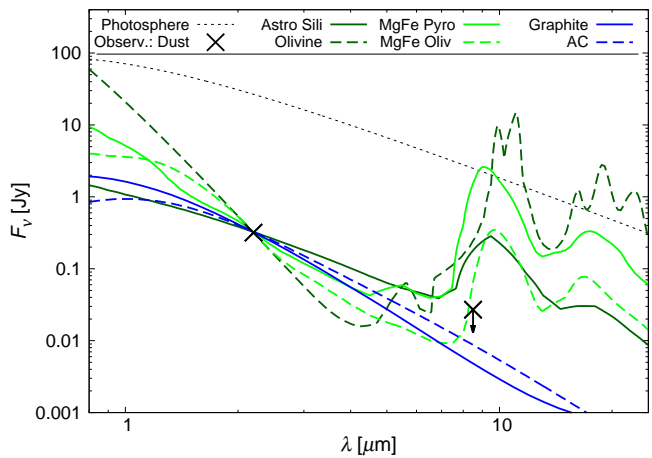}
\caption{SED considering different dust materials (see text) in a disc with ring radius $R=\unit[0.4]{au}$ and dust grains of size $a=\unit[0.1]{\text{\textmu}\textrm{m}}$ around HD~56537. The dust mass $M_{\mathrm{dust}}$ is chosen as a scaling factor to match the flux at $\lambda=\unit[2.2]{\text{\textmu}\textrm{m}}$ for each material. All silicates (green) show the characteristic spectral feature around $\lambda\sim\unit[10]{\text{\textmu}\textrm{m}}$ which causes an overestimation of the flux at $\unit[8.5]{\text{\textmu}\textrm{m}}$, contrary to the observations. In contrast, carbonaceous grains (blue) reproduce the observations.}
\label{Diff_Mat}
\end{figure} 
To investigate the impact of different dust materials on the disc modelling, appropriate simulations of the SED are performed for an exemplary disc with ring radius $R=\unit[0.4]{au}$ and dust grains of size $a=\unit[0.1]{\text{\textmu}\textrm{m}}$ (Fig.~\ref{Diff_Mat}). Besides graphite (optical data from \citealt{WeingartnerDraine2001}) from the previous section, we also consider:
\begin{itemize}
 \item Astronomical silicate (\textsl{Astro Sili};~\citealt{WeingartnerDraine2001}),
 \item Natural (terrestrial) olivine from San Carlos, Arizona (\textsl{Olivine}; data for 928 K; \citealt{Zeidler2011, Zeidler2015}),
 \item Glassy magnesium-iron silicates with pyroxene composition, Mg$_{0.5}$Fe$_{0.5}$Si$\,$O$_3$ (\textsl{MgFe~Pyro}; \citealt{Dorschner1995}),
 \item Glassy magnesium-iron silicates with olivine composition, Mg$\,$Fe$\,$Si$\,$O$_4$(\textsl{MgFe~Oliv}; \citealt{Dorschner1995}),
 \item Amorphous carbonaceous dust analogues produced by pyrolysis of cellulose at $\unit[1000]{{}^\circ C}$ (\textsl{AC}; \citealt{Jager1998}).
\end{itemize}
The first four materials are silicates and the last one is a carbon. The silicates show the characteristic broad spectral feature around $\lambda\sim\unit[10]{\text{\textmu}\textrm{m}}$. Regarding the thermal reemission as the dominating source of radiation, the flux at $\lambda=\unit[8.5]{\text{\textmu}\textrm{m}}$ is expected to be larger than the NIR~flux in most cases. However, the flux observed at $\lambda=\unit[8.5]{\text{\textmu}\textrm{m}}$ in systems with hot exozodiacal dust is one order of magnitude below the measured NIR~flux. We cannot reproduce the observational constraints under the assumption of dust grains consisting of any of the considered silicates. More specifically, only disc ring radii below the sublimation radius in combination with the dominance of large grains (where the spectral feature is less pronounced) would result in fluxes which are in agreement with the observations. Although the modelling cannot exclude the presence of large particles due to the limited resolution of the observation at  $\lambda=\unit[8.5]{\text{\textmu}\textrm{m}}$, we expect that they do not occur in the disc. Therefore, large amounts of silicates in the material of the exozodiacal dust are excluded.

In contrast, the crystalline graphite and the amorphous carbon can reproduce the observations, so that we focus on graphite in the following sections.
 
 
\subsection{Survey: Group~I-targets}
\label{survey_sd}
\begin{figure} 
 \includegraphics[trim=1.75cm 1.45cm 2.9cm 0.65cm, clip=true,width=1.0\linewidth, page=1]{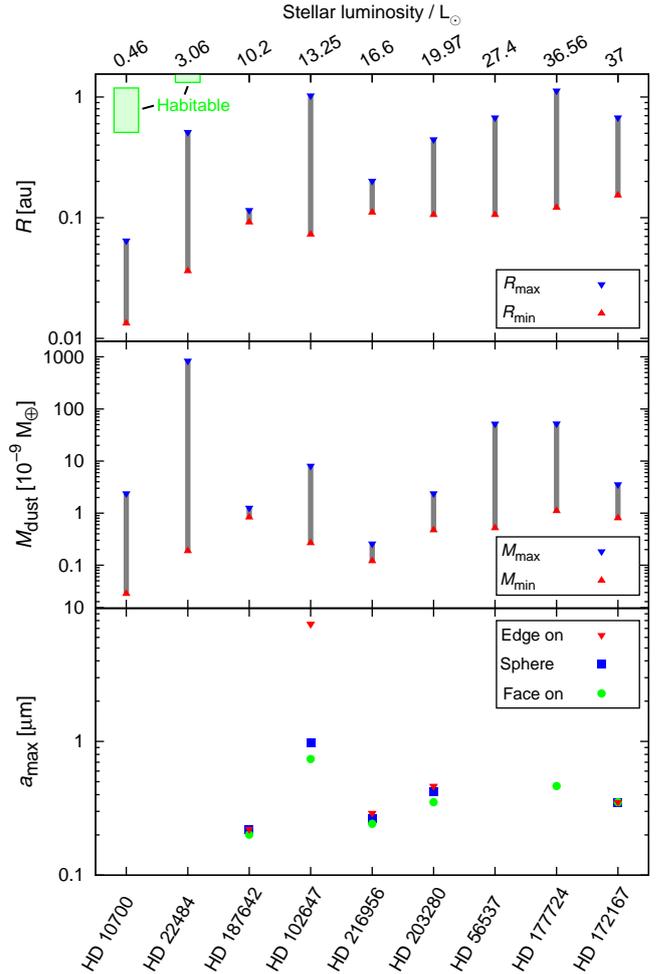} 
\caption{Results of the disc modelling of the Group~I-targets (Tab.~\ref{data_systems}). The stellar luminosity increases from left to right. 
\mbox{\textit{Top}: Minimum} and maximum disc ring radius, $R_\text{min}$ and $R_\text{max}$, within which the hot zodiacal dust is located. An approximate habitable zone between 210 and $\unit[320]{K}$ is marked for HD~10700 and HD~22484 as green boxes. For the other systems, the corresponding habitable zone is outside of the displayed radii.
\mbox{\textit{Middle}: Minimum} and maximum dust mass, $M_\text{min}$ and $M_\text{max}$, respectively. 
\textit{Bottom}: Maximum grain size for discs with face-on or edge-on orientation and for the spherical shell.}
\label{Result_all_Group1}
 \end{figure}

We now perform the calculations of the SED for each target of Group~I and determine the parameter settings which reproduce the observational data. The results are shown in Figure~\ref{Result_all_Group1}. 

The minimum disc ring radius $R_\text{min}$ varies between\break$\unit[\sim0.01]{au}$ and $\sim\unit[0.2]{au}$ and increases in a rough trend with the stellar luminosity. This is of course expected, since $R_\text{min}$ is in most cases determined by the sublimation radius $R_\text{sub}$.
 
The maximum disc ring radius $R_\text{max}$ is below $\unit[\sim 1]{au}$ for all systems, with a minimum value of $\unit[0.06]{au}$ for HD~10700. Interestingly, it also appears to increase with the stellar luminosity. In contrast to the minimum radius, this trend is not obvious, since we do not know which physical processes set the location of the exozodiacal dust and whether or not these are, for instance, temperature-dependent. Accordingly, a tentative trend might be important, and we now undertake a basic statistical analysis.

The Pearson correlation coefficient between $\log L_\star$ and $\log R_\text{max}$ amounts to $R_\text{p}=0.70$, and the Spearman rank correlation coefficient, which is more robust and does not require linearity, is $R_\text{s}=0.65$. If two apparent outliers, HD~22484 and HD~102647, are dropped from the analysis, the correlation gets even stronger ($R_\text{p}=0.85$, $R_\text{s}=0.94$), indicating a high significance.

\begin{figure}
 \includegraphics[trim=1.85cm 17.45cm 2.9cm 1.5cm, clip=true,width=1.0\linewidth, page=1]{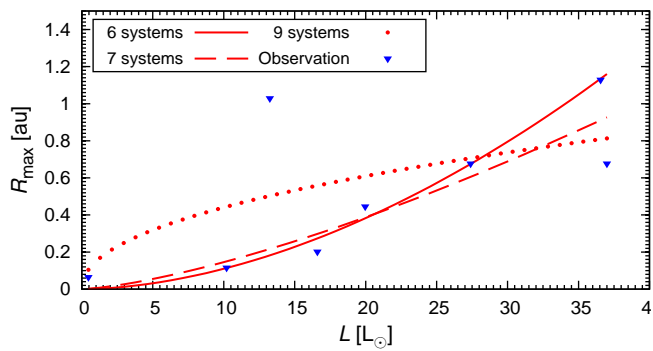} 
\vspace*{-0.4cm}
\caption{Maximum disc ring radius $R_\text{max}$ as a function of stellar luminosity $L_\star$ for the nine Group~I-targets (blue triangles), fitted with a function \mbox{$R_\text{max}=c_1\, L_\star^{c_{2}}$}. Excluding no system, or excluding HD 22484 and HD 102647, or excluding HD 22484, HD 102647 and Vega, the result is $R_\text{max} \propto L_\star^{0.5\pm 0.3}$ (dots), $R_\text{max} \propto L_\star^{1.4\pm 0.4}$ (dashed line), and $R_\text{max} \propto L_\star^{1.8\pm 0.2}$ (solid line), respectively.}
\label{Result_fit_Group1}
 \end{figure}

Since the trend is significant, we can make a power-law fit. The result strongly depends on how we treat the outliers (Fig.~\ref{Result_fit_Group1}). For the full sample of nine stars, i.e. not excluding any outliers, the result is \mbox{$R_\text{max} \propto L_\star^{0.5\pm 0.3}$}. Without two obvious outliers mentioned above we find \mbox{$R_\text{max} \propto L_\star^{1.4\pm 0.4}$}. Figure~\ref{Result_fit_Group1} suggests that HD~172167 (Vega) may be another outlier. Excluding all three systems results in \mbox{$R_\text{max} \propto L_\star^{1.8\pm 0.2}$}. Trying to understand which of these relations can be more trustable, we can look at all three potential outliers. Indeed, HD~22484 has the relatively small ratio of the measured NIR~flux to the flux at $\unit[8.5]{\text{\textmu}\textrm{m}}$, which results in weaker constraints for the model parameters (see Sect.~\ref{Indiv_one}). Thus we deem discarding this system reasonable. HD~102647 ($\beta$~Leo) is the only system with a significantly high flux at $\unit[8.5]{\text{\textmu}\textrm{m}}$, and this system reveals some differences to other resolved cases (see notes on this system in Sect.~\ref{Indiv_one}). As for Vega, there is some uncertainty about its luminosity (see Sect.~\ref{Indiv_one} for this star, too). Yet we do not see any justifiable reason for excluding $\beta$~Leo and Vega from the analysis. Discarding HD~22484 only, the best fit for the remaining eight systems is \mbox{$R_\text{max} \propto L_\star^{0.7\pm0.5}$}.  

Since we are interested in the most likely distance at which the hot dust is located, we may also choose to analyse the geometrical mean of the minimum and maximum radii, i.e. $R_\text{mean}=\sqrt{R_\text{min} R_\text{max}}$ instead of $R_\text{max}$. Doing this without the outlier HD~22848 leads to an even flatter dependence,
\mbox{$R_\text{mean} \propto L_\star^{0.6\pm0.2}$}. In fact, the dust location may scale as the square root of the stellar luminosity, which would imply that all exozodiacal dust discs have approximately the same temperature. If true, this could mean that the hot dust location is driven by temperature-dependent processes. Alternatively, this may be a chance coincidence. For instance, if the hot dust is trapped by stellar magnetic fields (see Sect.~\ref{401}), a similar trend may possibly arise if the Lorentz force scales with the stellar luminosity in a similar way. On any account, no definite conclusions can be made, given a small sample size and rather large uncertainties with which the dust ring radii are estimated.

In addition, we note that for all systems the location of the habitable zone defined in Sect.~\ref{sec_sublim} is outside of the location of the hot dust.

The range of grain sizes depends on whether the sublimation radius is resolved by the observation at $\lambda=\unit[8.5]{\text{\textmu}\textrm{m}}$ or not (HD~10700, HD~22484, HD~56537). If it is not, we are not able to put any constraints on the particle size. Since larger dust grains are not excluded in these systems, scattered radiation has a higher impact on the NIR~flux. HD~102647 and HD~177724 are marginally resolved by the MIR~observation, resulting in a large maximum grain radius $a_\text{max}$ for at least one of the three disc geometries. For the remaining systems of Group~I, the  maximum grain sizes are constrained to be between $\unit[0.2]{\text{\textmu}\textrm{m}}$ and $\unit[0.5]{\text{\textmu}\textrm{m}}$ and the deviations for different disc inclinations are below $\unit[10]{\%}$. Thermal reemission is the dominating source of the excess in the NIR. Consequently, the polarisation degree is below $\unit[5]{\%}$, and for grain sizes below $\unit[\sim0.2]{\text{\textmu}\textrm{m}}$ even below $\unit[1]{\%}$. The ratio of the scattered to total radiation amounts to values up to $\unit[35]{\%}$. However, the modelling results in a lower limit of the minimum particle size for the systems HD~10700, HD~187642 and HD~216956.

The maximum dust mass $M_\text{max}$ shows the same behaviour as the maximum disc ring radius $R_\text{max}$ and is in the range $\unit[(0.26-840)\times10^{-9}]{M_\oplus}$, while the minimum dust mass $M_\text{min}$ amounts to $\unit[(0.03-1.1)\times10^{-9}]{M_\oplus}$. Again, the maximum dust masses can be much better constrained for the systems where the sublimation radius is spatially resolved by the MIR~observation. In such systems, the maximum dust mass amounts to  $\unit[(0.2-3.5)\times10^{-9}]{M_\oplus}$.

\begin{figure}
 \includegraphics[trim=1.75cm 17.45cm 2.9cm 1.5cm, clip=true,width=1.0\linewidth, page=1]{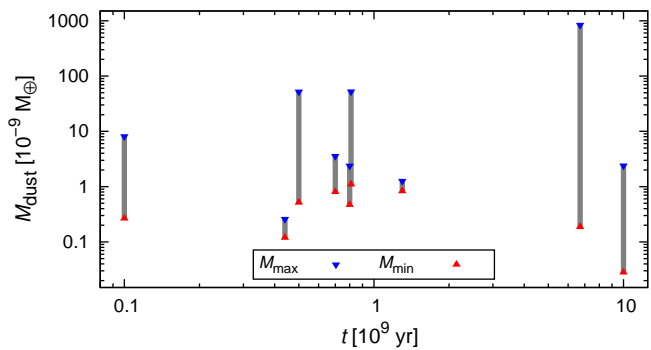} 
\vspace*{-0.4cm}
\caption{Minimum and maximum dust mass as a function of stellar age for the Group~I-targets (Tab.~\ref{data_systems}).}
\label{Result_tmass_Group1}
 \end{figure}

A decay of the dust mass with the stellar age is a robust theoretical prediction for ``classical'' debris discs, i.e., Kuiper-belt analogues (e.$\,$g. \citealt{Dominik2003, Wyatt2007, Loehne2008}). This trend has also been confirmed observationally, especially for A-type stars (e.$\,$g. \citealt{Rieke2005, Su2006}), although the dependence is much weaker for solar-type stars (e.$\,$g. \citealt{Eiroa2013, Montesinos2016}). However, it is not known whether the same is true for exozodiacal dust systems. To check this, we plotted the inferred minimum and maximum dust masses as a function of stellar age in Figure~\ref{Result_tmass_Group1}. With the Pearson correlation coefficient $R_\text{p}=0.29$ and the Spearman rank coefficient $R_\text{s}=0.11$ (neglecting of HD~22484 and HD~102647: $R_\text{p}=-0.10$, $R_\text{s}=0.04$), no strong correlation exists between the maximum mass of the hot exozodiacal dust and the age of the system.

\subsection{Individual targets of Group~I}
\label{Indiv_one}
In this section, individual sources of Group~I are discussed briefly.
 \begin{figure*}
\vspace*{-0.2cm} \includegraphics[trim=2.1cm 12.4cm 2.4cm 12.11cm, clip=true,width=1.01\linewidth, page=1]{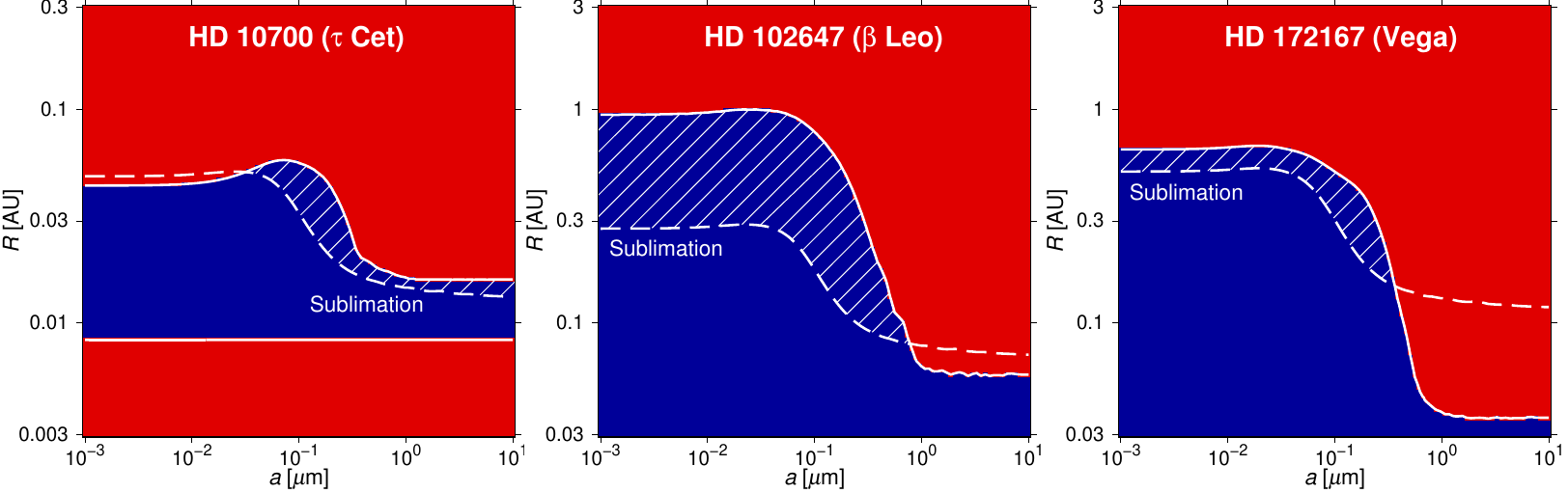}\vspace*{-0.2cm}
\caption{Same as Figure~\ref{Para1a}, but for the systems HD~10700 ($\tau$~Cet), HD~102647 ($\beta$~Leo), and HD~172167 (Vega), and only for the face-on case. 
The dust material is pure graphite. For HD~10700, the sublimation radius and the upper blue-red border constrain the minimum dust grain size, while for HD~102647 and HD~172167 the the maximum grain radii are constrained. HD~102647 is the only system of the sample with a significant excess at $\unit[8.5]{\text{\textmu}\textrm{m}}$ (\citealt{Mennesson2014}).}
\label{Para2a} 
 \end{figure*}

\subsubsection*{HD~10700 ($\tau$~Cet)}
HD~10700 is one of the brightest sun-like stars in our neighbourhood (G8V, $K=1.7^\text{mag}$) and with $\sim\unit[10^{10}]{yr}$ the oldest of the sample. The K band excess was first reported by \cite{diFolco2007}. While there is no significant excess at $\lambda=\unit[8.5]{\text{\textmu}\textrm{m}}$ or $\unit[24]{\text{\textmu}\textrm{m}}$, \cite{Habing2001} reported excess at $\unit[60]{\text{\textmu}\textrm{m}}$ and $\unit[100]{\text{\textmu}\textrm{m}}$ and 
\cite{Greaves2004} and \cite{Lawler2014} presented resolved images at $\unit[70]{\text{\textmu}\textrm{m}}$, $\unit[160]{\text{\textmu}\textrm{m}}$ and $\unit[850]{\text{\textmu}\textrm{m}}$.

We find that the hot dust has to be located between $\unit[0.013]{au}$ and $\unit[0.06]{au}$. The minimum dust mass is about $\unit[0.03\times10^{-9}]{M_\oplus}$. The maximum dust mass takes values from $\unit[1.2\times10^{-9}]{M_\oplus}$ (face-on) to $\unit[2.4\times10^{-9}]{M_\oplus}$ (edge-on). Since the sublimation radius of the large particles is resolved for the NIR~observation but not for the MIR~observation, the maximum grain size is not constrained. Particles smaller than $\unit[1]{\text{\textmu}\textrm{m}}$ sublimate at larger distances than larger particles. For the face-on disc and the spherical dust sphere of HD~10700, this condition sets a lower limit for the grain sizes (see Fig.~\ref{Para2a}), and the minimum particle radius is $\unit[34]{\textrm{nm}}$ and $\unit[22]{\textrm{nm}}$, respectively.

\subsubsection*{HD~22484 (10~Tau)}
\cite{Kospal2009} showed that the SED exhibits an excess at $\unit[70]{\text{\textmu}\textrm{m}}$ for HD~22484, while there is no significant excess at $\unit[24]{\text{\textmu}\textrm{m}}$. The ratio of the measured NIR~flux to the flux at $\unit[8.5]{\text{\textmu}\textrm{m}}$ is relatively small ($\sim3.5$). 

The scattered radiation contributes to the NIR~flux up to $\unit[\sim30]{\%}$ (for grains between $\unit[0.1]{\text{\textmu}\textrm{m}}$ and $\unit[1]{\text{\textmu}\textrm{m}}$), even though thermal reemission is the dominant source for the NIR~flux at a wide range of the suitable parameter settings. The disc ring radii  are not strongly affected by the disc inclination and take values between $\unit[0.035]{au}$ and $\unit[0.52]{au}$. The minimum dust mass is about  $\unit[0.2\times10^{-9}]{M_\oplus}$ and the maximum dust mass is in the range $\unit[(150-840)\times10^{-9}]{M_\oplus}$. The grain sizes are not constrained.

\subsubsection*{HD~56537 ($\lambda$~Gem)}
The system was discussed in detail in Section~\ref{represent}. Besides the K band excess, there was no significant signal above the photosphere at $\unit[8.5]{\text{\textmu}\textrm{m}}$, $\unit[24]{\text{\textmu}\textrm{m}}$ or $\unit[70]{\text{\textmu}\textrm{m}}$ (\citealt{Gaspar2013, Chen2014, Mennesson2014}).

\subsubsection*{HD~102647 ($\beta$~Leo, Denebola)}
The K band excess of HD~102647 was first reported by \cite{Akeson2009}. It is the only system of the sample with a significant excess at $\unit[8.5]{\text{\textmu}\textrm{m}}$ (\citealt{Mennesson2014}). HD~102647 also has a significant excess at both $\unit[24]{\text{\textmu}\textrm{m}}$ and $\unit[70]{\text{\textmu}\textrm{m}}$ (\citealt{Su2006}) and is well studied (e.$\,$g. \citealt{Stock2010, Churcher2011}), revealing a relative compact disc at $\unit[60-70]{au}$ and with further dust components located close to $\unit[2]{au}$ to the star. \text{Herschel} images show a nearly face-on disc (\citealt{Matthews2010}).

In our disc modelling we derived the location of the hot exozodiacal dust to be between $\unit[0.07]{au}$ and $\unit[1]{au}$. The minimum dust mass amounts to $\unit[0.3\times10^{-9}]{M_\oplus}$. The maximum dust mass takes values from $\unit[5.8\times10^{-9}]{M_\oplus}$ (face-on) to $\unit[8.1\times10^{-9}]{M_\oplus}$ (edge-on). Since the sublimation radius is resolved for all particle sizes by the NIR and also marginally by the MIR~observation, the maximum grain size is constrained to $\unit[0.74]{\text{\textmu}\textrm{m}}$ (face-on, Fig.~\ref{Para2a}),  $\unit[7.6]{\text{\textmu}\textrm{m}}$ (edge-on) and  $\unit[1.0]{\text{\textmu}\textrm{m}}$ (sphere).

\subsubsection*{HD~172167 ($\alpha$~Lyr, Vega)}
HD~172167 shows excess in H~band (\citealt{Defrere2011}) and K~band (\citealt{Absil2006}), while the flux at $\unit[8.5]{\text{\textmu}\textrm{m}}$ is not significant but only slightly below the significance limit (\citealt{Mennesson2014}). \cite{Su2006} reported a large excess at $\unit[70]{\text{\textmu}\textrm{m}}$, which corresponds to a disc nearly face-on (\citealt{Monnier2012}). Vega is a rapid rotator (\citealt{Peterson2006}) which makes stellar parameters a function of the stellar latitude. Consequently, the luminosity at the equator and at the poles derived from equatorial and polar values of the stellar radius $R_\star$ and temperature $T_\star$ deviate between $\unit[7900-10150]{K}$ and $\unit[28-57]{L_\odot}$, respectively (\citealt{AufdenBerg2006}). In our simulations, we adopt intermediate values of $T_\star=\unit[9620]{K}$ and $\unit[37]{L_\odot}$ (\citealt{Mueller2010}). 

We find that the dust has to be located between $\unit[0.15]{au}$ and $\unit[0.68]{au}$ and the dust mass amounts to $\unit[(0.8-3.6)\times10^{-9}]{M_\oplus}$. The sublimation radius is resolved for all particle sizes by both the NIR and the MIR~observation. Consequently, the maximum grain size is well constrained\break ($<\unit[0.35]{\text{\textmu}\textrm{m}}$, Fig.~\ref{Para2a}). These values are comparable to the parameters found by \cite{Absil2006} and \cite{Defrere2011}, who indicated grains of sizes smaller than $\unit[0.2]{\text{\textmu}\textrm{m}}$ and distances of $R\unit[\sim0.2-0.3]{au}$.

\subsubsection*{HD~177724 ($\zeta$~Cep), HD~187642 ($\alpha$~Aql, Altair), and~HD~203280~($\alpha$~Cep, Alderamin)}
These three systems are all showing an excess in the K~band but are free of excess at any longer wavelengths (\citealt{Habing2001, Chen2005, Plavchan2009, Gaspar2013, Mennesson2014}).

For HD~177724, the minimum  and maximum of the inner disc ring radius amount to about $\unit[0.12]{au}$ and $\unit[1.1]{au}$, respectively. The minimum mass is between $\unit[1.1\times10^{-9}]{M_\oplus}$ and $\unit[1.3\times10^{-9}]{M_\oplus}$ and the maximum mass lies between $\unit[10\times10^{-9}]{M_\oplus}$ and $\unit[52\times10^{-9}]{M_\oplus}$. For the edge-on case and the spherical dust shell, the sublimation radius is unresolved for the large dust grains, and thus the maximum dust grain size not constrained. For the face-on case, the maximum grain size amounts to $\unit[0.46]{\text{\textmu}\textrm{m}}$.

For HD~187644, the dust is strongly confined to a range between $\unit[0.09]{au}$ and $\unit[0.11]{au}$ and the mass amounts to $\unit[(0.8-1.3)\times10^{-9}]{M_\oplus}$. The minimum grain size  is about $\unit[0.15]{\text{\textmu}\textrm{m}}$ and the maximum grain size amounts to $\unit[0.22]{\text{\textmu}\textrm{m}}$.

For HD~203280, the inner dust ring radius is located between $\unit[0.1]{au}$ and $\unit[0.45]{au}$ with a dust mass of $\unit[(0.47-2.4)\times10^{-9}]{M_\oplus}$. The maximum grain size is constrained to $\unit[0.35]{\text{\textmu}\textrm{m}}$ (face-on),  $\unit[0.46]{\text{\textmu}\textrm{m}}$ (edge-on) and  $\unit[0.42]{\text{\textmu}\textrm{m}}$ (sphere).

\subsubsection*{HD~216956 ($\alpha$~PsA, Fomalhaut)}
The well-studied debris disc system of Fomalhaut shows excess in K~band (\citealt{Absil2009}), while the excess at $\unit[8.5]{\text{\textmu}\textrm{m}}$ is not significant (\citealt{Mennesson2014}). At slightly longer wavelengths, an unresolved emission was detected with Spitzer/IRS which is interpreted as warm dust (\citealt{Stapelfeldt2004}; \citealt{Su2013}). At $\unit[70]{\text{\textmu}\textrm{m}}$ and $\unit[850]{\text{\textmu}\textrm{m}}$ a large excess is observed which is assigned to reemission of the cold debris belt at $\unit[140]{au}$ (\citealt{Acke2012}; \citealt{Boley2012}).

In our disc modelling we derived the minimum and maximum disc ring radii of the hot exozodiacal dust to $\unit[0.11]{au}$ and $\unit[0.2]{au}$, respectively. The dust mass amounts to $\unit[(0.12-0.26)\times10^{-9}]{M_\oplus}$ and the minimum grain size to $\unit[0.1]{\text{\textmu}\textrm{m}}$. Since the sublimation radius is resolved for all particle sizes by both the NIR and the MIR~observation, the maximum grain size is constrained to $\unit[0.24]{\text{\textmu}\textrm{m}}$ (face-on),  $\unit[0.29]{\text{\textmu}\textrm{m}}$ (edge-on) and  $\unit[0.27]{\text{\textmu}\textrm{m}}$ (sphere). These values are comparable to the parameters found in the modelling of \cite{Lebreton2013}, where $a<\unit[0.5]{\text{\textmu}\textrm{m}}$ and $R\unit[\sim0.1-0.3]{au}$.
\subsection{Survey: Group~II-targets}
\label{survey_se}
\begin{figure} 
 \includegraphics[trim=1.7cm 8.6cm 2.9cm 0.75cm, clip=true,width=1.0\linewidth, page=1]{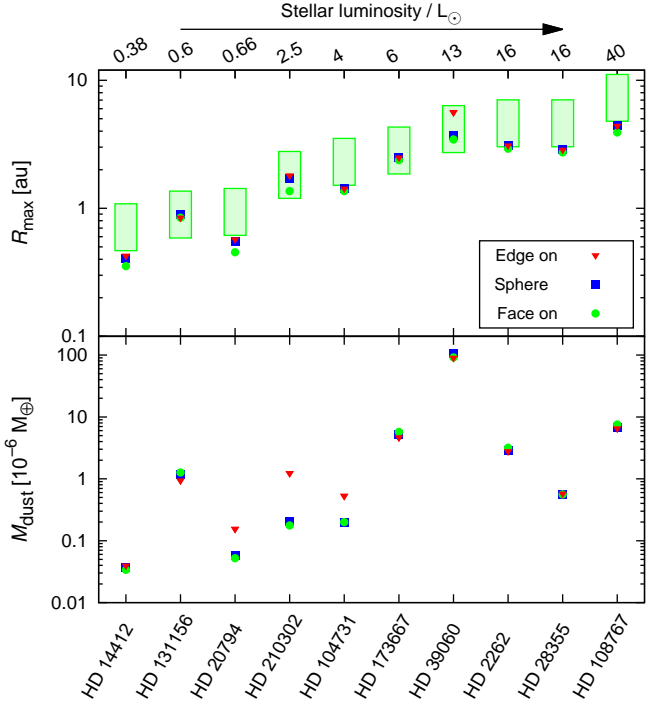} 
\caption{Results of the disc modelling of the Group~II-targets (Tab.~\ref{data_systems}) for discs with face-on or edge-on orientation and for the spherical shell. The stellar luminosity increases from left to right. 
\textit{Top}: Maximum disc ring radius $R_\text{max}$, within which the hot zodiacal dust is located. Approximate habitable zones between 210 and $\unit[320]{K}$ are marked as green boxes for each system.
\textit{Bottom}: Maximum dust mass.}
\label{Result_all_Group2}
 \end{figure}

Targets of Group~II are those systems with a circumstellar NIR~excess for which no interferometric observations at wavelengths  $\lambda=\unit[8.5]{\text{\textmu}\textrm{m}}$ exist. We performed the calculations of the SED also for each target of Group~II and determined the parameter settings which reproduce the observational data. The results are shown in Figure~\ref{Result_all_Group2}. 

The grain size could not be constrained for any system of Group~II, i.$\,$e. grain sizes from $\unit[1]{nm}$ to $\unit[100]{\text{\textmu}\textrm{m}}$ all reproduce the observed NIR~excess without overestimating fluxes in the MIR or FIR. 

The maximum ring radii $R_\text{max}$ which are in agreement with the observational constraints amount to values between $\unit[0.3]{au}$ and $\unit[6]{au}$, i.$\,$e. they are a factor of six larger than in the case of Group~I-targets. As a result, in five of the ten systems dust is potentially located in the habitable zone (while in the other five it is close to the inner edge of that zone). The dust masses amount to up to $\unit[1\times10^{-4}]{M_\oplus}$ and are thus more than two magnitudes larger than the dust masses of the Group~I-targets. Of course, this does not mean that the hot discs of Group~II are more massive and larger than the discs of Group~I. The only reason for the difference is that interferometric observations around $\sim\unit[10]{\text{\textmu}\textrm{m}}$, available for the Group~I, constrain the disc parameters much more tightly. 

Similar to the Group I-targets, the maximum disc ring radius $R_\text{max}$ seems to increase with increasing stellar luminosity. Only HD~39060 ($\beta$~Pic) contradicts to this trend, for which a large disc ring radius up to $\sim \unit[6]{au}$ is inferred. This is a consequence of large MIR and FIR~fluxes above the stellar photosphere, which are caused by dust at larger radii of the massive disc around $\beta$~Pic. Moreover, $\beta$~Pic with an age of $\unit[23\pm3]{Myr}$ (\citealt{Mamajek2014}) may still be in the process of planet formation, especially in the inner $\unit[10]{au}$ of the disc. Thus, the observed NIR~excess could be related to the end of the planet building phase rather than to exozodiacal dust (\citealt{Absil2008}). 

The Pearson and the Spearman rank correlation coefficients between $\log L_\star$ and $\log R_\text{max}$ of all targets of Group~II amount $R_\text{p}=0.94$ and $R_\text{s}=0.91$, respectively (neglecting $\beta$~Pic: $R_\text{p}=0.97$ and $R_\text{s}=0.96$), indicating a significant trend. Nevertheless, it should be noted that for the Group~II-targets only the observed fluxes at $\lambda=\unit[24]{\text{\textmu}\textrm{m}}$ and $\unit[70]{\text{\textmu}\textrm{m}}$ are used which are not cleaned by the contribution of the stellar photosphere (Sect.~\ref{sec_data}), so the derived correlation between $\log L_\star$ and $\log R_\text{max}$ could be caused by the contribution of the stellar photosphere rather than by the emission of the dust.

While the weak dust emission on top of the bright stellar emission at wavelengths around $\sim\unit[10]{\text{\textmu}\textrm{m}}$ is difficult to deduce from the photometry, interferometric observations provide a direct estimation of the flux ratio between the circumstellar environment and the photosphere of the star. Well determined fluxes at these wavelengths are necessary to find better constraints for the parameters of the hot exozodiacal dust. Therefore, high resolution interferometric observations around $\unit[\sim10]{\text{\textmu}\textrm{m}}$ of the Group~II-targets are required. HD~108767 and HD~173667 will be targets of the planned survey ``Hunt for Observable Signatures of Terrestrial planetary Systems (HOSTS)'' on the Large Binocular Telescope Interferometer (\citealt{Weinberger2015}), observing in the N$^\prime$ band ($\lambda=\unit[9.8-12.4]{\text{\textmu}\textrm{m}}$).

\subsection{Uncertainty of the NIR~excess}
\label{sec_uncertainty}
This study intends to put general constraints on the hot dust, based on the observational data available.
An essential source of information is the NIR data. However, the NIR~excess measurements have large $1\,\sigma$-uncertainties, typically between $\unit[10]{\%}$ and $\unit[30]{\%}$. Thus it is important to check how the constraints we derive would change if the actual NIR~flux were larger or smaller by $1\,\sigma$ than the nominal value.

We do this again for the exemplary system HD~56537, for which the NIR~flux is $\unit[27]{mJy}\pm\unit[5.8]{mJy}$. The uncertainty of this flux measurement is $\unit[21]{\%}$. Assuming the flux is increased (decreased) by $1\,\sigma$ results in a decreasing (increasing) of the maximum disc ring radius of $\unit[20]{\%}$ for the three disc inclinations. Since the sublimation radius is still the minimum disc ring radius, $R_\text{min}$ and the dust masses are unaltered.

\subsection{Variability of the NIR~excess}
\label{sec_time_variab}
Using VLTI/PIONIER in H band, \cite{Ertel2016} performed follow-up observations for seven Group~II-targets. They
concluded that the phenomenon responsible for the excesses persists for the majority of the systems for the explored time scales of at least two to four years. Six of the seven systems show no significant variation, while HD~7788 is the first strong candidate for variability. Therefore, HD~7788 was excluded from our study.

For the three remaining targets of Group~II as well as for the complete Group~I no follow-up observations exist yet, and no specific statements can be made concerning the variability of the excess of these individual targets. However, considering the fact that \cite{Ertel2016} found at least one of seven systems to show significant variability, the assumption of a temporal constant NIR~excess in our study might not hold for all our targets. As shown in Section~\ref{sec_uncertainty}, this might have extensive influence on the dust parameters derived for these individual systems. For example, if we assumed the NIR~excess of HD~56537 to by a factor of 2 lower, the maximum disc ring radius would increase by a factor of $1.9$. Only further follow-up observations can reveal time variability or constancy of the excess in these systems.

Besides the stability of the NIR detections over at least a few years, it is still unclear if the hot dust exists for longer than ten years. \cite{Ertel2014b} found a detection rate of hot exozodiacal dust in the H band of $\unit[11]{\%}$, so if the dust has lifetimes below ten years, it has to be replenished in time intervals smaller than $\unit[\sim100]{years}$. As long as the formation mechanisms of the hot dust are unknown, this can be clarified again only by follow-up observations over a longer period. If the phenomenon responsible for the excesses is stable for time scales of ten to twenty years, it is highly probable that the excess of most systems already existed in the past when they were observed with Spitzer, which has been up to eight years before the NIR~observations. If this is not the case, new observations in the MIR would be necessary to clarify if the phenomenon responsible for the NIR~excess causes significant excess at wavelengths beyond $\unit[20]{\text{\textmu}\textrm{m}}$. This would influence primarily the results of the Group~II-targets, since the modelling parameters of these targets are constrained by the Spitzer observations, while the results of the Group~I-targets are mainly constrained by the interferometric observations at $\lambda=\unit[8.5]{\text{\textmu}\textrm{m}}$, which have been taken up to four years before the NIR~observations.

In summary, only follow-up and quasi-simultaneous multi-wavelength observations can clarify the time variability or stability of the hot dust phenomenon.


\subsection{Spectrally dispersed data}
\label{sec_disp}
The PIONIER data come from three spectral channels across the H band ($\lambda=\unit[1.59]{\text{\textmu}\textrm{m}}, \unit[1.68]{\text{\textmu}\textrm{m}}, \unit[1.77]{\text{\textmu}\textrm{m}}$) which allows one to investigate the spectral slope of the excess of Group~II-targets. These spectrally dispersed data may allow one to put further constraints on the location and the sizes of the dust grains, though the significance of the slopes is quite small (\citealt{Defrere2012, Ertel2014b}).

We apply the fitted slopes $a_\text{slope}$ and their uncertainties $\sigma_\text{slope}$ as derived by \cite{Ertel2014b}. Among all the systems of Group~II, HD~108767 has the smallest uncertainty with $\sigma_\text{slope}=1.35$. For each parameter setting in Table~\ref{free_parameter}, we calculate the fluxes at $\lambda=\unit[1.59]{\text{\textmu}\textrm{m}}$ and $\unit[1.77]{\text{\textmu}\textrm{m}}$ for HD~108767 and determine the spectral slope of the simulated data. As suggested by \cite{Ertel2014b}, a $3\,\sigma_\text{slope}$ criterion is used to prove if the measured slope coincides with the simulated slopes within the uncertainties.

Unfortunately, the spectrally dispersed data could not further constrain the parameters. Almost the entire parameter space considered would yield spectral slopes in the H band which are consistent within the uncertainties with the simulated slopes ($a_\text{slope}\in\left[-2.58,\,5.52\right]$). In particular, the suitable parameter settings as derived from the sublimation radius and the ratio of NIR to MIR fluxes result in slopes below $5$. Stronger constraints from the spectral slope require data of higher precision or over a wider wavelength range to minimise the uncertainties of the slopes (e.$\,$g.~using~\text{VLTI/MATISSE}).

\section{Discussion}
\label{401}
Our disc modelling shows that grains have to be absorbing (e.$\,$g. composed of graphite), have to be located within $\unit[1]{au}$, and have to be smaller than $\unit[0.5]{\text{\textmu}\textrm{m}}$ in radius (at least for systems with spatially resolved sublimation radius in the MIR). However, the disc models presented here only show which properties the dust must have to reproduce the observational data, without addressing the physical mechanisms responsible for the production and evolution of that dust in the system.

Explaining the presence of small grains in significant amounts close to the star poses a challenge (e.$\,$g.~\citealt{Matthews2014}). If this dust was produced in a steady-state collisional cascade operating in an asteroid belt analogue, such a belt would rapidly deplete. The resulting dust fractional luminosities would be several orders below the observed ones, at least in Gyr-old systems (e.$\,$g. \citealt{Wyatt2007a,Loehne2008,Ertel2014b}).

An additional problem arises from the fact that the dust grains are inferred to be small enough to be either swiftly expelled from the system by direct radiation pressure or to undergo a rapid inward drift to the sublimation region by the Poynting-Robertson force.
The radiation blow-out and Poynting-Robertson timescale (\citealt{Burns1979}) of $\unit[1]{\text{\textmu}\textrm{m}}$ sized grains at $\unit[1]{au}$ is on the order of $\sim10^3$ years for the lowest luminosity star of the sample ($L_\star=\unit[0.46]{L_\odot}$) and even below $\sim10^2$ years for the highest luminosity star ($L_\star=\unit[37]{L_\odot}$). If a local dust production in a steady-state regime is assumed, these short timescales will imply unrealistically high dust replenishment rates.

In view of these difficulties, alternative scenarios have been proposed.
These include transient dynamical events similar to the late heavy bombardment in the solar system (\citealt{Wyatt2007a, Kennedy2013}), the sublimation of a ``supercomet" close to the star (\citealt{Beichman2005}), or the aftermath of a large collision (\citealt{Lisse2008, Lisse2009, Jackson2014}). However, these scenarios either fail to explain the presence of exozodiacal dust around old stars
or have a low probability, which is incompatible with the observed frequency of systems with hot dust. Another scenario is the delivery of material from exterior debris belts to the star's vicinity, e.$\,$g. by comets dynamically perturbed from the outer cold discs, planetesimals or planets and disintegrating in the inner system (e.$\,$g.~\citealt{Faramaz2017}). However, this scenario suffers from a number of rather unconstrained parameters such as unknown number and parameters of scattering planets and unknown efficiency of comet disintegration (\citealt{Bonsor2012, Bonsor2013}). Yet another possibility, an inward drift of dust grains by the Poynting-Robertson drag with subsequent grain sublimation, would predict a much higher flux in the MIR than is actually observed (\citealt{Kobayashi2009, Kobayashi2011}). 
 
Currently, the most promising model is that of charged, nano-scale dust grains which are trapped in the stellar magnetic field through the Lorentz force (\citealt{Czechowski2010, Su2013, Rieke2016}). Such a model can be further improved by including a more realistic sublimation physics for dust grains (\citealt{Lebreton2013}). For example, MgO, FeO and C nano-grains have melting temperatures larger than $\unit[3000]{K}$, allowing grains of these or composite materials to survive closer to the star. Nano-particles can be charged through the photoelectric effect (for early-type stars) or stellar wind (for late-type stars; \citealt{Pedersen2011}). With the assumption of surface fields similar to the Sun ($\unit[\sim1]{G}$), the magnetic fields will then be able to cage the nano-grains in the systems, causing the measured NIR~fluxes.

\section{Conclusions}
\label{501}
Based on near- to far-infrared observations of a sample of nine systems harbouring hot exozodiacal dust, we derived the following constraints on the dust properties and dust distribution in the vicinity of these stars:
\begin{itemize}
\item The observed fluxes in the NIR and at a wavelength of $\lambda=\unit[8.5]{\text{\textmu}\textrm{m}}$ are inconsistent with dust compositions dominated by silicate material. Instead, carbonaceous materials such as graphite are compatible with the observations.
\item The maximum disc ring radius $R_\text{max}$ amounts to $\unit[0.06]{au}$ to $\unit[1]{au}$. We find evidence for $R_\text{max}$ to increase with the stellar luminosity $L_\star$. It is possible, albeit unsure, that $R_\text{max}$ scales as a square root of $L_\star$, which would imply a nearly constant temperature of hot dust in all the systems studied.

\item The maximum grain radius is constrained to  $\unit[0.2-0.5]{\text{\textmu}\textrm{m}}$ for systems for which the sublimation radius is spatially resolved by the interferometric MIR~observation.  If it is not, the particle size is not or poorly constrained by our study. The result is independent of whether the observed interferometric MIR~flux is significant or not.

\item For some of these systems, we are also able to constrain the minimum grain size. To be consistent with the observed fluxes and to lie outside the sublimation distance, the dust grains have to be larger that $\unit[20-150]{\textrm{nm}}$ in radius.

\item The dust masses in the systems are derived to range between $\unit[0.2\times10^{-9}]{M_\oplus}$ and $\unit[3.5\times10^{-9}]{M_\oplus}$ for systems for which the sublimation radius is spatially resolved by the interferometric MIR~observation. If it is not, the masses range between $\unit[0.26\times10^{-9}]{M_\oplus}$ and $\unit[840\times10^{-9}]{M_\oplus}$. There is no obvious correlation between the dust mass and the age of the system.

\item Thermal reemission is the dominant source of the excess not only in the MIR and FIR, but also in the NIR. Nevertheless, for systems whose
sublimation radius is not spatially resolved by the interferometric MIR~observation, a contribution of scattered light of up to $\unit[35]{\%}$ cannot be completely excluded.

\item The polarisation degree of the NIR radiation of the dust is below $\unit[5]{\%}$ for all systems, and for grains smaller than $\unit[0.2]{\text{\textmu}m}$ even below $\unit[1]{\%}$. This finding is consistent with recently published polarisation observations.
\end{itemize}
We also modelled another ten systems known to harbour exozodiacal dust, for which $\lambda=\unit[8.5]{\text{\textmu}\textrm{m}}$ data are not available. The results for these systems appear similar, albeit with weaker constraints on the grain size, dust location, and dust masses.

The origin of the exozodiacal dust is still a matter of debate. Models that consider charged nano-grains trapped in stellar magnetic fields, combined with a proper choice of materials and realistic sublimation physics, can yield a satisfactory explanation for the existence of the exozodiacal dust. The results of this paper pose tighter constraints on the hot dust material and may help improve and verify such models.

\section*{Acknowledgements}
The authors thank the DFG for financial support under contracts WO 857/13-1, WO 857/15-1, KR~2164/15-1, and HM~1164/9-1. We also thank Bertrand Mennesson for useful information on the fluxes of interferometric nulling data and the anonymous referee for a precious help in improving the quality of the paper.

 \bibliographystyle{mnras}
{\footnotesize
 \bibliography{Literature}
}
\label{lastpage}

\bsp	

\end{document}